\documentclass[sigconf]{acmart}
\usepackage{xcolor}
\usepackage{amsmath}
\usepackage{multirow}
\usepackage{array}
\usepackage{tabularx}
\usepackage{hyperref}
\usepackage[linesnumbered,ruled,vlined]{algorithm2e}
\usepackage[noend]{algpseudocode}
\usepackage{placeins}
\usepackage{colortbl}
\usepackage{siunitx}

\AtBeginDocument{%
  }

\copyrightyear{2023}
\acmYear{2023}
\setcopyright{rightsretained}
\acmConference[KDD '23]{Proceedings of the 29th ACM SIGKDD Conference on Knowledge Discovery and Data Mining}{August 6--10, 2023}{Long Beach, CA, USA}
\acmBooktitle{Proceedings of the 29th ACM SIGKDD Conference on Knowledge
Discovery and Data Mining (KDD '23), August 6--10, 2023, Long Beach, CA, USA}
\acmDOI{10.1145/3580305.3599831}
\acmISBN{979-8-4007-0103-0/23/08}

\newcommand{\cov}{\mathrm{Cov}}
\newcommand{\var}{\mathrm{Var}}
\DeclareMathOperator*{\argmin}{argmin}
\DeclareMathOperator*{\argmax}{argmax}
\newcommand{\parens}[1]{\left(#1\right)}
\newcommand{\brackets}[1]{\left[#1\right]}
\newcommand{\braces}[1]{\left\lbrace #1\right\rbrace}
\newcommand{\angles}[1]{\left\langle #1\right\rangle}
\newcommand{\Verts}[1]{\left\Vert #1\right\Vert}
\newcommand{\expectation}[2]{\mathbb{E}_{#1}\left[#2\right]}

\SetKwInput{KwInput}{Input}
\SetKwInput{KwOutput}{Output} 

\definecolor{n0040}{RGB}{180, 235, 170}
\definecolor{n0035}{RGB}{183, 232, 167}
\definecolor{n0030}{RGB}{186, 230, 164}
\definecolor{n0025}{RGB}{190, 228, 161}
\definecolor{n0020}{RGB}{193, 226, 159}
\definecolor{n0015}{RGB}{196, 224, 156}
\definecolor{n0010}{RGB}{200, 222, 153}
\definecolor{n0005}{RGB}{203, 220, 151}
\definecolor{p0000}{RGB}{206, 218, 148}
\definecolor{p0005}{RGB}{210, 216, 145}
\definecolor{p0010}{RGB}{213, 214, 143}
\definecolor{p0015}{RGB}{216, 212, 140}
\definecolor{p0020}{RGB}{220, 210, 137}
\definecolor{p0025}{RGB}{223, 208, 134}
\definecolor{p0030}{RGB}{226, 206, 132}
\definecolor{p0035}{RGB}{230, 204, 129}
\definecolor{p0040}{RGB}{233, 202, 126}
\definecolor{p0045}{RGB}{236, 200, 124}
\definecolor{p0050}{RGB}{225, 170, 89}
\definecolor{p0055}{RGB}{243, 196, 118}
\definecolor{p0060}{RGB}{243, 178, 74}

\begin{document}

\title{Generating Synergistic Formulaic Alpha Collections via Reinforcement Learning}

\begin{abstract}
    In the field of quantitative trading, it is common practice to transform raw historical stock data into indicative signals for the market trend. Such signals are called alpha factors. Alphas in formula forms are more interpretable and thus favored by practitioners concerned with risk. In practice, a set of formulaic alphas is often used together for better modeling precision, so we need to find synergistic formulaic alpha sets that work well together. However, most traditional alpha generators mine alphas one by one separately, overlooking the fact that the alphas would be combined later. In this paper, we propose a new alpha-mining framework that prioritizes mining a synergistic set of alphas, i.e., it directly uses the performance of the downstream combination model to optimize the alpha generator. Our framework also leverages the strong exploratory capabilities of reinforcement learning~(RL) to better explore the vast search space of formulaic alphas. The contribution to the combination models' performance is assigned to be the return used in the RL process, driving the alpha generator to find better alphas that improve upon the current set. Experimental evaluations on real-world stock market data demonstrate both the effectiveness and the efficiency of our framework for stock trend forecasting. The investment simulation results show that our framework is able to achieve higher returns compared to previous approaches.
\end{abstract}

\ccsdesc[500]{Computing methodologies~Reinforcement learning}
\ccsdesc[300]{Computing methodologies~Search methodologies}
\ccsdesc[300]{Applied computing~Economics}

\keywords{Computational Finance, Stock Trend Forecasting, Reinforcement Learning}


\author{Shuo Yu}
\authornote{These authors contributed equally.}
\authornote{Key Lab of Intelligent Information Processing of Chinese Academy of Sciences (CAS). Xiang Ao is also at Institute
of Intelligent Computing Technology, Suzhou, China.}
\author{Hongyan Xue}
\authornotemark[1]
\authornotemark[2]
\author{Xiang Ao}
\authornotemark[2]
\authornote{Corresponding authors.}
\affiliation{%
    \institution{Institute of Computing Technology, Chinese Academy of Sciences}
    \institution{University of Chinese Academy of Sciences}
    \city{Beijing}
    \country{China}
}
\email{yushuo19b@ict.ac.cn}
\email{xuehongyan21b@ict.ac.cn}
\email{aoxiang@ict.ac.cn}

\author{Feiyang Pan}
\author{Jia He}
\author{Dandan Tu}
\affiliation{%
    \institution{Huawei EI Innovation Lab}
    \country{China}
}
\email{pfy824@gmail.com}
\email{hejia0149@gmail.com}
\email{tudandan@huawei.com}

\author{Qing He}
\authornotemark[2]
\authornotemark[3]
\affiliation{%
    \institution{Institute of Computing Technology, Chinese Academy of Sciences}
    \institution{University of Chinese Academy of Sciences}
    \city{Beijing}
    \country{China}
}
\email{heqing@ict.ac.cn}

\maketitle

\section{Introduction}

Currently, it is almost a standard paradigm to transform raw historical stock data into indicative signals for the market trend in the field of quantitative trading~\cite{qian2007quantitative}.
These signal patterns are called \emph{alpha factors}, or {alphas} in short~\cite{tulchinsky15}. Discovering alphas with high returns has been a trendy topic among investors and researchers due to the close relatedness between alphas and investment revenues.

The prevailing methods of discovering alphas can be in general divided into two groups, namely machine learning-based and formulaic alphas. Most recent research has focused on the former ones. These more sophisticated alphas are often obtained via deep learning models, e.g., using sequential models like LSTM~\cite{lstm}, or more complex ones integrating non-standard data like HIST~\cite{hist} and REST~\cite{rest}, etc. 
On the other end of the spectrum, we have the alphas that can be represented in simple formula forms. Such formulaic alphas are traditionally constructed by human experts using their domain knowledge and experience, often expressing clear economic principles. To name some, \cite{alpha101} demonstrates 101 alpha factors tested on the US stock market. Recently, research has also been conducted on frameworks that generate such formulaic alphas automatically \cite{alphaevolve,autoalpha,huataigp,huataigp2}. These approaches are able to find loads of new alphas rapidly without human supervision, while still maintaining relatively high interpretability compared to the more sophisticated machine learning-alphas.

Despite the existing approaches achieving remarkable success, however, they still have disadvantages in different aspects. 
Machine learning-based alpha factors are inherently complex and sometimes require more complex data other than the price/volume features.
In addition, although they are often more expressive, they often suffer from relatively low explainability and interpretability. 
As a result, when the performance of these ``black box'' models unexpectedly deteriorates, it is hard for human experts to tune the models accordingly.
These algorithms are thus not favored under some circumstances due to concerns about risks.
On the other hand, while formulaic alphas are more interpretable, previous research on this matter often focused on finding a single alpha factor that predicts well on its own. 
Nonetheless, it is often impossible to describe a complex and chaotic system such as the stock market with simple rules that human researchers can comprehend.
As a compromise, a set of these alphas are oftentimes used together in practice, instead of using them individually.
However, when multiple of these independently mined formula alphas are combined, the final prediction performance may not improve much because not much consideration is put into the synergistic effect between factors~(see Section \ref{sec:ablation} for detail). 
In addition, these alphas are often simple in their forms, and their underlying mechanisms are often quite understandable. 
Once they are released to the public and become well-known among practitioners, their performance may deteriorate rapidly~\cite{alpha101}. 

Therefore, the question we are facing is: \emph{Are we able to find a way to automatically discover interpretable alpha factors, which work well with downstream predictive models, without suffering possible performance deterioration due to the alpha factors being widely known to the general public?}

To solve the above challenge, we formulate a new research problem in this paper, which is to find \emph{synergistic formulaic alpha factor sets}. Using raw stock price/volume data as the input, we aim to search for a set of formulaic alpha factors instead of individual ones. 
Recall that finding a single well-performing alpha on given data is already a hard problem to resolve since the search space of valid formulas is vast and hard to navigate. 
The search space for alpha mining is often even larger than that of a typical symbolic regression problem~\cite{dsr}. 

The most intuitive approach to this problem would be using genetic programming~(GP), performing mutations on expression trees to generate new alphas. In fact, most previous work on this matter is based on genetic programming~(GP) \cite{huataigp,huataigp2,alphaevolve,autoalpha}, which is of course not a serendipitous choice since GP methods generally excel at such problems with large search spaces. 
However, GP algorithms often scale poorly due to the complexity of maintaining and mutating a huge population~\cite{dsr}. In addition, the main challenge remains that mining a set of synergistic alphas all at once is an even harder problem with a much larger search space, the scale of which makes most existing frameworks infeasible to solve. 

Hence, previous works mostly tried to find ways to simplify the problem of alpha set mining, by mining alphas one by one and filtering out a subset of them with respect to some similarity metric. The mutual information coefficient~(IC) between the pairs of alpha in the set is often employed as the similarity ``metric''~\cite{alphaevolve,autoalpha,huataigp}. However, as we will demonstrate below, adding a new alpha that is of high IC to the ones in an existing pool of alpha may still bring a non-negligible boost of performance to the combined result, and vice versa. This phenomenon still exists even when the combination model is set to be a simple linear regressor. Therefore, the traditional approach to determining whether a set of alpha could be synergistic does not line up with the expected outcome.

To tackle the challenge that GP methods could be inefficient at exploring the vast search space of formulaic alphas, our framework utilizes reinforcement learning~(RL) for achieving better results in exploration. 
Combined with the strong expressiveness of deep neural networks, RL with its excellent exploratory ability plays a predominant role in numerous areas. To list a few examples, game playing~\cite{muzero}, natural language processing~\cite{rlnlp}, symbolic optimization~\cite{dsr}, and portfolio management~\cite{deeptrader}. 
We implement a sequence generator with constraints to ensure valid formulaic alpha generation and employ a policy gradient-based algorithm to train the generator in the absence of a direct gradient.
Since traditional mutual-IC filtering methods do not align well with the target of optimizing the combination model's performance, we propose to use directly the performance as the optimization objective of our alpha generator.
Under this new optimization scheme, our generator is able to produce a synergistic set of alpha which fits the mine-and-combine procedure in a more suitable way. 
To evaluate our alpha-mining framework, we conduct extensive experiments over real-world stock data. Our experiment results demonstrate that the formulaic alpha sets generated by our framework perform better than those generated with previous approaches, shown both on the prediction metrics and investment simulations.

Our contributions can be summarized as follows.
\begin{itemize}
    \item We propose a new optimization scheme that produces a set of alpha that suits downstream tasks better, regardless of what actual form the combination model takes.
    \item We introduce a new framework for searching formulaic alpha factors based on policy gradient algorithms, to utilize the strong exploratory power of reinforcement learning.
    \item We present a series of experimental results demonstrating the effectiveness of our proposed framework. Additional experiments and case studies are also conducted to demonstrate why mutual IC-based filtering techniques that are previously commonly used may not work as expected when considering the combined performance of an alpha set.
\end{itemize}

\section{Problem Formulation}\label{sec:formulation}

\subsection{Alpha Factor}

We consider a stock market with $n$ stocks in a period of $T$ trading days in total. On each trading day $t \in \braces{1, 2, \cdots, T}$, each stock $i$ corresponds to a feature vector $x_{ti} \in \mathbb{R}^{m\tau}$, comprised of $m$ raw features such as opening/closing price in the recent $\tau$ days\footnote{This ``unrolling'' of historical data introduces redundancy. Namely, feature vectors of a stock on consecutive trading days have overlapping sections. This notation is chosen for the convenience of demonstration.}. Finally, we define an \emph{alpha factor} $f$ as a function mapping feature vectors of all stocks on a trading day $X \in \mathbb{R}^{n\times m\tau}$ into \emph{alpha values} $z = f(X) \in \mathbb{R}^n$. We will use the word ``alpha'' for both an alpha factor and its corresponding values in the following sections.

\subsection{Alpha Factor Mining}

To measure the effectiveness of an alpha, we calculate the information coefficient (IC) between the true stock trend it aims to predict $y_t \in \mathbb{R}^n$ and the factor values $f(X_t)$. We denote the daily IC function as $\sigma: \mathbb{R}^{n} \times \mathbb{R}^{n} \to [-1, 1]$, which is defined as the Pearson's correlation coefficient:
\begin{equation}
    \sigma(u_t, v_t) = \frac{\sum_{i=1}^{n} (u_{ti} - \bar{u}_t)(v_{ti} - \bar{v}_t)}{\sqrt{\sum_{i=1}^{n} (u_{ti} - \bar{u_t})^2 \sum_{i=1}^{n} (v_{ti} - \bar{v_t})^2}}.
    \label{eq:pearson}
\end{equation}
Such value can be calculated on every trading day between an alpha and the prediction target. For convenience, we denote the IC values between two sets of vectors averaged over all trading days as $\bar{\sigma}(u, v) = \expectation{t}{\sigma(u_t, v_t)}$.

We use the average IC between an alpha and the return to measure the effectiveness of an alpha factor on a stock trend series $y = \braces{y_1, y_2, \cdots, y_T}$:
\begin{equation}
    \bar{\sigma}_y(f) = \bar{\sigma}(f(X), y).
\end{equation}

As mentioned above, the output of a combination model can be seen as a ``mega-alpha'', mapping raw inputs into alpha values. Therefore, we denote the combination model as $c(X;\mathcal{F},\theta)$, where $\mathcal{F} = \braces{f_1, f_2, \cdots, f_k}$ is a set of alphas to combine, and $\theta$ denotes the parameters of the combination model. We would like the combination model to be optimal w.r.t. a given alpha set $\mathcal{F}$ on the training dataset $y$, that is:

\begin{equation}
    \begin{aligned}
        c^*(X; \mathcal{F}) & = c(X; \mathcal{F}, \theta^*), \quad \textrm{where} \\
        \theta^* & = \argmax_\theta \bar{\sigma}_y(c(\cdot; \mathcal{F}, \theta)).
    \end{aligned}
\end{equation}

Conclusively, the task of mining a set of alphas can be defined as the optimization problem $\argmax_\mathcal{F} c^*(\cdot; \mathcal{F})$.

\begin{figure}[t]
    \centering
    \includegraphics[width=0.7\linewidth]{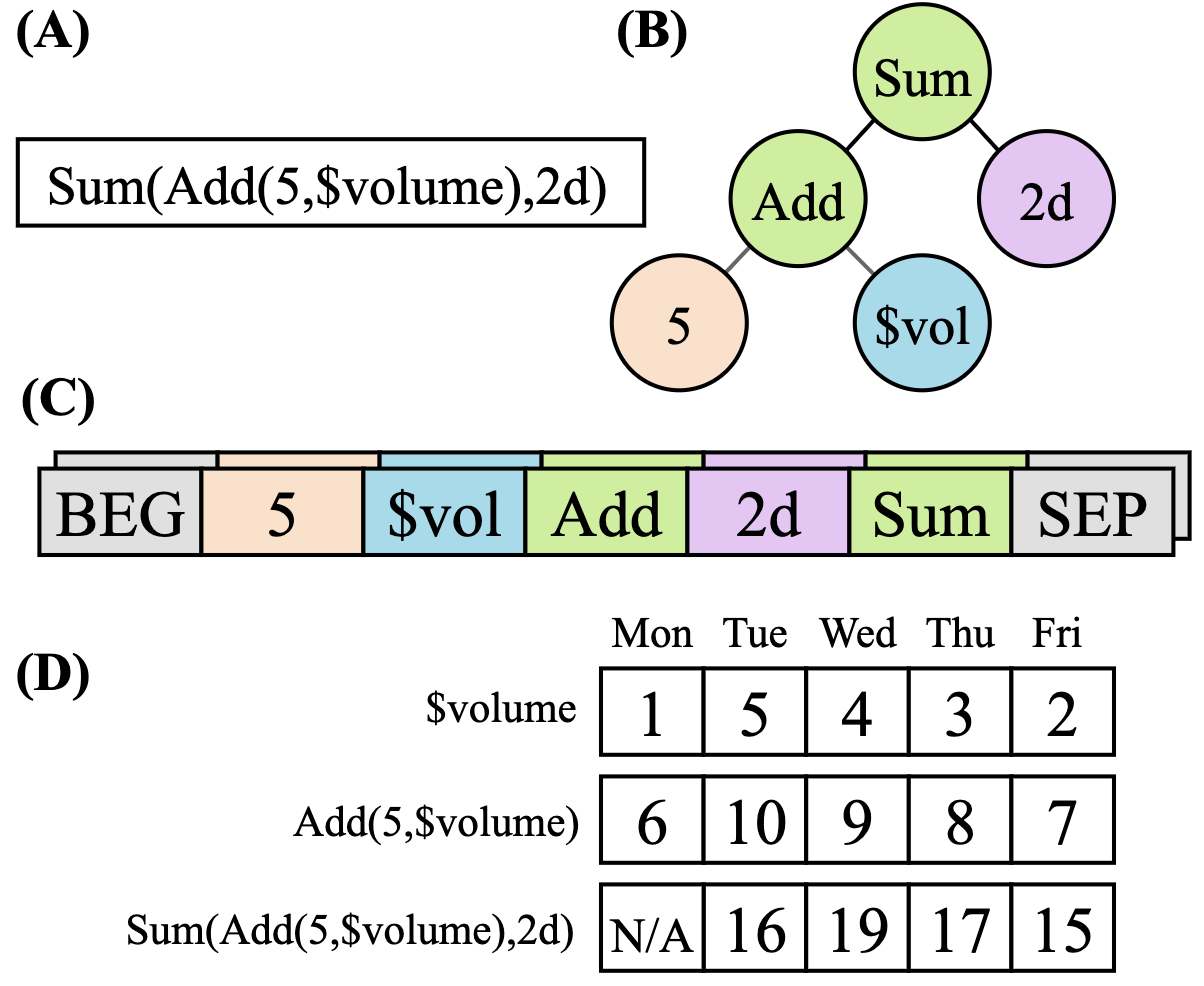}
    \caption{(A) An example of a formulaic alpha. (B) Its equivalent expression tree. (C) Its reverse Polish notation (RPN). Note that \emph{BEG} and \emph{SEP} are sequence indicators later mentioned in our framework. (D). Step-by-step computation of this alpha on an example time series.}
    \label{fig:formulation}
\end{figure}

\subsection{Formulaic Alpha}

Formulaic alphas are expressed as mathematical expressions, consisting of various operators and the raw input features mentioned before. Some examples of the operators are the elementary functions (like ``$+$'' and ``$\log$'') operated on one-day data, called cross-section operators, and operators that require data from a series of days, called time-series operators (e.g. ``Min(close, 5)'' gives the lowest closing price of a stock in the recent 5 days). A list of all the operators used in our framework is given in Appendix \ref{sec:ops}.

Such formulas can be naturally represented by an expression tree, with each non-leaf node representing an operator, and children of a node representing the operands. To generate such an expression, our model represents the expression tree by its postorder traverse, with the children's order also defined by the traversing order. In other words, the model represents a formula as its reverse Polish notation (RPN). It is easy to see that such notation is unambiguous since the arities of the operators are all known constants. See Figure \ref{fig:formulation} for an example of a formulaic alpha expression together with its corresponding tree and RPN representations.

\section{Methodology}\label{sec:method}

\begin{figure*}[!tphb]
    \centering
    \includegraphics[width=0.8\linewidth]{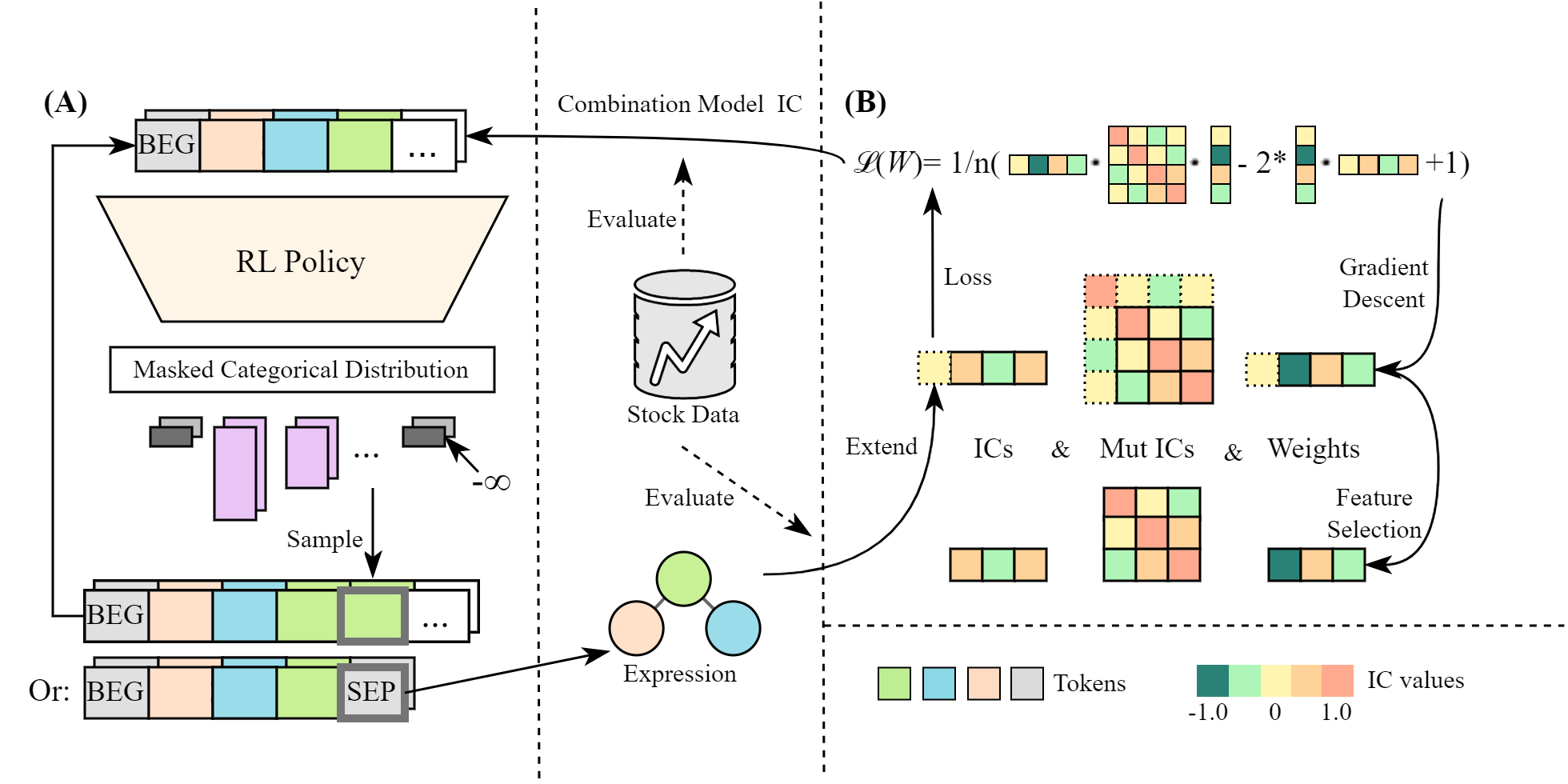}
    \caption{An overview of our alpha-mining framework. (A) An alpha generator that generates expressions, optimized via a policy gradient algorithm. (B) A combination model that maintains a weighted combination of principal factors and, in the meantime, provides evaluative signals to guide the generator.}
    \Description{}
    \label{fig:main}
\end{figure*}

As illustrated in Figure \ref{fig:main}, our alpha-mining framework consists of two main components: 1) the \emph{Alpha Combination Model}, which combines multiple formulaic alphas to achieve optimal performance in prediction, and 2) the \emph{RL-based Alpha Generator}, which generates formulaic alphas in the form of a token sequence. The performance of the Alpha Combination Model is used as the reward signal to train the RL policy in the Alpha Generator using policy gradient-based algorithms, such as PPO \cite{ppo}. Repeating this process, the generator is continuously trained to generate alphas that boost the combination model, thereby enhancing the overall predictive power.

\subsection{Alpha Combination Model}  \label{sec:alpha_improvement}

Considering the interpretability of the combined ``mega-alpha'', the combination model itself should also be interpretable. In this paper, we use a linear model to combine the alphas.

The values evaluated from different alphas have drastically different scales, which might cause problems in the following optimization steps. To counter this effect, we centralize and normalize the alpha values with their average and standard deviation. Since Pearson's correlation coefficient is invariant up to linear transformation, this transformation does not affect the performance of the alphas when they are considered separately. Formally, we introduce a normalization operator $\mathcal{N}$, that transforms a vector such that its elements have a mean of 0, and the vector has a length of 1:
\begin{equation}
    \brackets{\mathcal{N}(u)}_i = \frac{u_i - \bar{u}}
    {\sqrt{\sum_{j=1}^n\parens{u_j - \bar{u}}^2}}.
\end{equation}

We will omit explicitly writing the $\mathcal{N}$ operator for simplicity. For the rest of this paper, we will assume that all the $f(X)$ evaluations and the targets $y$ are normalized to have a mean of 0 and a length of 1 before subsequent computations. In other words, treat $f$ as $\mathcal{N} \circ f$ and $y$ as $\mathcal{N}(y)$.

Given a set of $k$ alpha factors $\mathcal{F} = \braces{f_1, f_2, \cdots, f_k}$ and their weights $w = \parens{w_1, w_2, \cdots, w_k} \in \mathbb{R}^k$, the combination model $c$ is defined as follows:
\begin{equation}
    c(X; \mathcal{F}, w)
    = \sum_{j=1}^{k} w_j f_j(X) = z.
\end{equation}

We define the loss of the combination model as the mean squared error (MSE) between model outputs and true stock trend values:
\begin{equation}
    \mathcal{L}(w) = \frac{1}{nT}\sum_{t=1}^{T} \|z_t - y_t\|^2.
\end{equation}

To simplify the calculation of alpha combination, we have:

\begin{theorem} \label{thm:loss_transformation}
Let $\mathcal{F}$ be a set of $k$ alphas and $w$ be their respective weights, the MSE loss $\mathcal{L}(w)$ can be represented as:

\begin{equation} \label{eq:model_loss}
    \begin{aligned}
        \mathcal{L}(w) = \frac{1}{n} \parens{
            1 - 2 \sum_{i=1}^{k} w_i \bar{\sigma}_y(f_i) + \sum_{i=1}^{k} \sum_{j=1}^{k} w_i w_j \bar{\sigma}(f_i(X), f_j(X))
        }.
    \end{aligned}
\end{equation}

\end{theorem}

The proof of this theorem is provided in Appendix \ref{sec:thm_proof}. Notice that there is no $z_t$ term on the RHS of Equation \ref{eq:model_loss}. Once we have obtained $\bar{\sigma}_{y}(f)$ for each alpha $f$ and their pairwise mutual correlations $\bar{\sigma}(f_i(X),f_j(X))$, we can then calculate the loss $\mathcal{L}(w)$ solely using these terms, saving time on calculating the relatively large $z_t$ in each gradient descent step.
 
Considering time and space complexity, it is impractical to combine all generated alphas together, because to calculate mutual correlation for each pair of factors we need $\mathcal{O}(k^2)$ evaluations of mutual IC. The quadratic growth of this makes it expensive to apply the current procedure to a large number of alphas. However, a few dozen of alphas will suffice for practical uses. To a certain point, more alphas would not bring much more increment in performance, following the law of diminishing returns. We will demonstrate this effect in Section \ref{sec:ablation}.

\begin{algorithm}
\caption{Incremental Combination Model Optimization}
\label{alg:model}
\KwInput {Alpha set $\mathcal{F} = \braces{f_1, \cdots, f_k}$, weights $w = \braces{w_1, \cdots, w_k}$ and a new alpha $f_\textrm{new}$}
\KwOutput {Optimal alpha subset $\mathcal{F}^* = \braces{f_1', \cdots, f_k'}$, optimal weights $w^* = \parens{w_1', \cdots, w_k'}$}
$\mathcal{F}\gets \mathcal{F}\cup \braces{f_\textrm{new}}, w\gets w \Vert \mathrm{rand}()$;\\
\ForEach{$f \in \mathcal{F}$}{
    Obtain $\bar{\sigma}_y(f)$ from calculation or cache;\\
    \ForEach{$f' \in \mathcal{F}$}{
        Obtain $\bar{\sigma}(f(X), f'(X))$ from calculation or cache;
    }
}
\For{$i\gets1$ \KwTo $num\_gradient\_steps$}{
    Calculate $\mathcal{L}(w)$ according to Equation \ref{eq:model_loss};
    $w \gets \mathrm{GradientDescent}(\mathcal{L}(w))$;
}
$p \gets \argmin_i |w_i|$;\\
$\mathcal{F} \gets \mathcal{F}\backslash \braces{f_p}$, $w \gets \parens{w_1, \cdots, w_{p-1}, w_{p+1}, \cdots, w_k}$;\\
\Return $\mathcal{F}, w$;
\end{algorithm}

After the alpha generator outputs a new alpha, the alpha is first added to the candidate alpha set and assigned a random initial weight. Gradient descent is then performed to optimize the weights with respect to the extended alpha set. We also set a threshold to limit the size of the alpha set, leaving only the principal alphas with the largest absolute weight. If the amount of alphas in the extended set exceeds a certain threshold, the least principal alpha is removed from the set together with its corresponding weight. The pseudocode of the training procedure is shown in Algorithm \ref{alg:model}.

\subsection{Alpha Generator} \label{sec:alpha_generation}

\begin{figure}[h]
  \centering
  \includegraphics[width=1.0\linewidth]{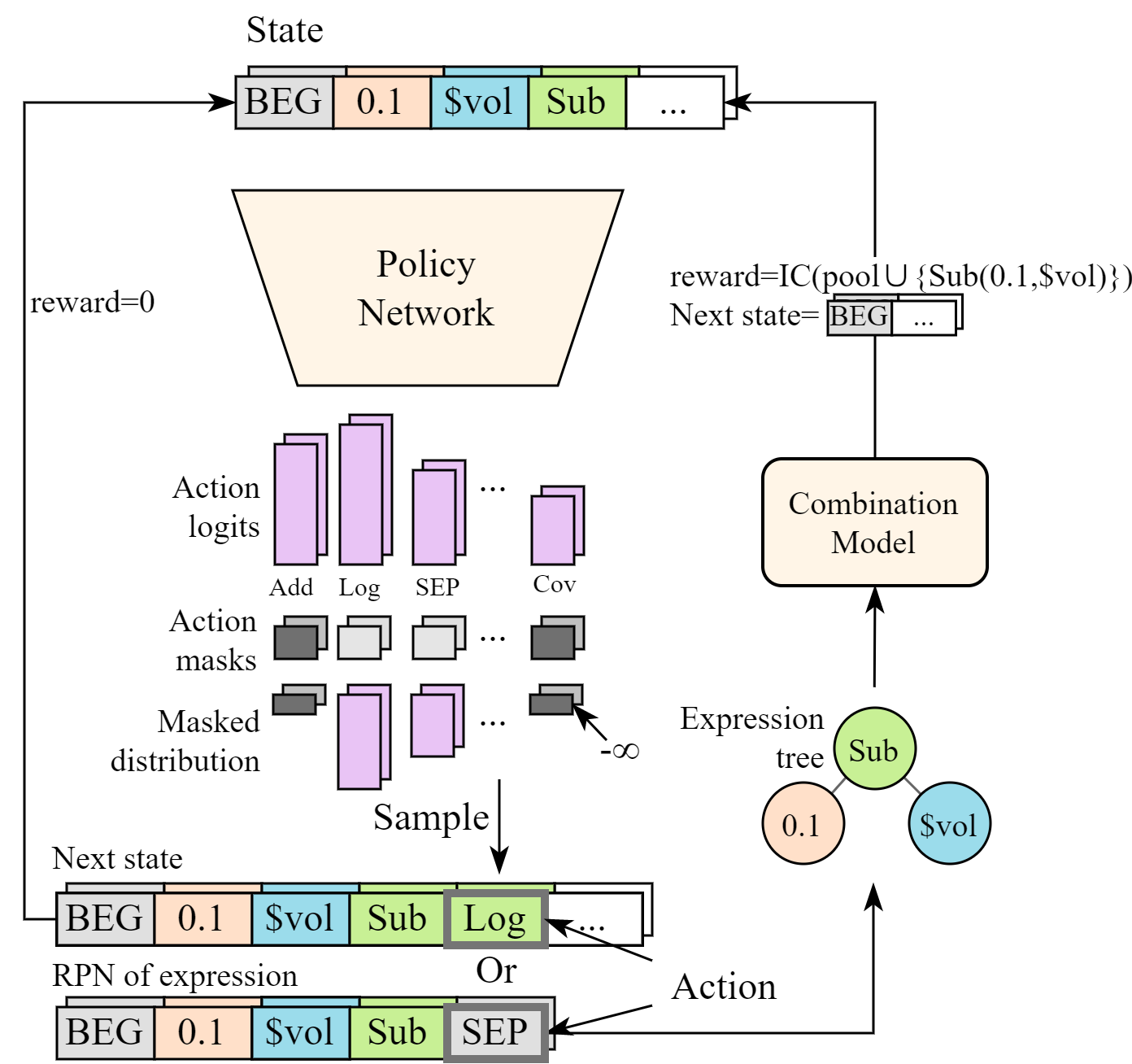}
  \caption{An illustration of our alpha generation framework.}
  \label{fig:rl}
\end{figure}

The alpha generator models a distribution of mathematical expressions. As each expression can be represented as a symbolic expression tree, we use the reverse Polish notation (RPN) to represent it as a linear sequence, since traditional auto-regressive generators can only deal with sequences. To control and evaluate the generation process of valid expressions, we model the generation process as a non-stationary Markov Decision Process (MDP). We will describe the various components of the MDP below in the following paragraphs. An overview of the MDP-based Alpha generator is shown in Figure \ref{fig:rl}.

\subsubsection{Tokens}

The token is an important abstraction in our framework. A token can be any of the operators, the features, or constant values. Table \ref{tab:tokens} shows some examples of such tokens. For the full list of operators, please refer to Section \ref{sec:ops}; for the full list of features we have chosen, please refer to Section \ref{sec:data}.

\begin{table}[!t]
	\centering
	\caption{Tokens used in our framework.}
	\resizebox{0.48\textwidth}{!}{
		\begin{tabular}{c|c}
			\toprule[1.5pt]
			Category & Examples \\
			\midrule[1pt]
			Operators & \emph{CS-Log}, \emph{CS-Add}, \emph{TS-Mean}, \emph, $\dotsc$ \\
            Features & \emph{\$open}, \emph{\$volume}, $\dotsc$ \\
			Constants & $-30, -10, -5, -2, -1, -0.5, -0.01, 0.01, 0.5, 1, 2, 5, 10, 30$ \\
			Time Deltas & $10d, 20d, 30d, 40d, 50d$ \\
			Sequence Indicator & \emph{BEG}(begin), \emph{SEP}(end of expression) \\
			\bottomrule[1.5pt]
	\end{tabular}}
	\label{tab:tokens}
\end{table}

\subsubsection{State Space}

Each state in the MDP corresponds to a sequence of tokens denoting the currently generated part of the expression. The initial state is always \emph{BEG}, so a valid state always starts with \emph{BEG} and is followed by previously chosen tokens. Since we aim for interpretability of the alphas, and too long of a formula will instead be less interpretable, we cap the length threshold of the formulas at 20 tokens.

\subsubsection{Action Space} \label{sec:action_space}

An action is a token that follows the current state (generated partial sequence). It is obvious that an arbitrarily generated sequence is not guaranteed to be the RPN of an expression, so we only allow a subset of actions to be taken at a specific state to guarantee the well-formedness of the RPN sequence. Please refer to Appendix \ref{sec:expr_legality} for more details.

\subsubsection{Dynamics}

Given a state and an action, we can obtain the next state deterministically. The next state is generated by taking the current state's corresponding sequence and appending the action token at the end.

\subsubsection{Rewards and Returns}

The MDP does not give immediate rewards for partially formed sequences. At the end of each episode, if the final state is valid, the state will be parsed to a formulaic function and evaluated in the combination model shown in Algorithm \ref{alg:model}. To encourage our generator to generate novel alphas, we will then evaluate the new combination model with the new alpha added, and use the model's performance as the return of this episode. Since the reward varies together with the components of the alpha pool, the MDP is non-stationary.

Contrary to common RL task settings, for alpha expression generation we do not necessarily want to penalize longer episodes (longer expressions). In fact, longer alphas that perform well are harder to find than shorter ones, due to exponential explosion of the search space. Consequently, 
we set the discount factor as $\gamma = 1$ (no discount).

\begin{algorithm}
\caption{Alpha Mining Pipeline}
\label{alg:rl}
\KwInput {Stock trend dataset $Y = \braces{y_t}$}
\KwOutput {Optimal alpha subset $F^* = \braces{f_1', \cdots, f_k'}$, optimal weights $w^* = \braces{w_1', \cdots, w_k'}$}
Initialize $\mathcal{F}$ and $w$;\\
Initialize RL policy $\pi_{\theta}$ with parameters $\theta$ and replay buffer $\mathcal{D}$;\\
\For{each iteration}{
    \For{each environment step}{
        $a_t \sim \pi_{\theta}(a_t | s_t)$;\\
        $s_{t+1} \gets [s_t, a_t]$;\\
        \eIf {$a_t = \emph{SEP}\ or\ len(s_{t+1}) \ge threshold$}{
            $f \gets parse(s_{t+1})$;\\
            Update $\mathcal{F}$, $w$ using $f$ and Algorithm \ref{alg:model};\\
            $IC_{new} \gets \bar{\sigma}_y(\sum_{i=1}^{k} w_i f_i)$;\\
            $r_t \gets IC_{new}$;
        }{
            $r_t \gets 0$;
        }
        $\mathcal{D} \gets \mathcal{D} \cup \braces{(s_t, a_t, r_t, s_{t+1})}$;
    }
    \For{each gradient step}{
        Use batch $\mathcal{B} \subset \mathcal{D}$ to do gradient descent on PPO objective $\mathcal{L}^{CLIP}(\theta)$ to update $\theta$;
    }
}
\Return $\mathcal{F}, w$;
\end{algorithm}

\subsubsection{Reinforcement Algorithm}

Based on the MDP defined above, we use Proximal Policy Optimization (PPO) \cite{ppo} to optimize a policy $\pi_\theta(a_t | s_t)$ that takes a state as input and outputs a distribution of action. An actual action will be sampled from the output distribution.

PPO is an on-policy RL algorithm based on the trust region method. It proposed a clipped objective $\mathcal{L}^{CLIP}$ as follows:

\begin{equation}
    \mathcal{L}^{CLIP}(\theta) = \hat{\mathbb{E}}_{t}\brackets{\min\braces{r_t(\theta)\hat{A}_{t},\mathrm{clip}\parens{r_t(\theta), 1-\epsilon, 1+\epsilon}\hat{A}_{t}}},
\end{equation}
where $r_t(\theta)=\frac{\pi_\theta(a_t | s_t)}{\pi_{old}(a_t | s_t)}$ and $\hat{A}_t$ is an estimator of the advantage function at timestep $t$. Using the importance sampling mechanism, PPO can effectively take the biggest possible improvement while keeping the policy in a trust region that avoids accidental performance collapse.

Since our MDP has complicated rules for the legality of actions, an action sampled from the full discrete action distribution predicted by the learned policy is likely to be invalid as mentioned in Section \ref{sec:action_space}. We adopt the Invalid Action Masking mechanism \cite{ppomask} to mask out invalid actions and just sample from the set of valid actions.

\subsection{Network Architecture}

The PPO algorithm requires the agent to have a value network and a policy network. Under our experiment settings, the two networks share a base LSTM feature extractor that converts token sequences into dense vector representations. Separate value and policy ``heads'' are attached after the LSTM. The values of hyperparameters are given in Appendix \ref{sec:hyperparam}.

\subsection{Training with policy gradient-based methods}

For the task of alpha mining, we do not require the agent to achieve relatively high average returns in each episode, but place more importance on the trajectories the agent takes in the whole training process. For this reason, we maintain a pool of alphas without resetting between episodes. We run the alpha generation procedure mentioned in Section \ref{sec:alpha_generation} and optimize the alpha combination model according to Section \ref{sec:alpha_improvement} repeatedly. In this way, we train the policy to continuously generate novel alpha factors that bring improvement to the overall prediction performance.

The proposed alpha mining process is shown in Algorithm \ref{alg:rl}. Our implementation is publicly available\footnote{ \url{https://github.com/RL-MLDM/alphagen/}}.

\section{Experiments}\label{sec:exp}

Our experiments are designed to investigate the following questions:

\begin{itemize}
    \item \textbf{Q1}: How does our proposed framework compare to prior alpha mining methods?
    \item \textbf{Q2}: How well does our model scale as the alpha set size increases?
    \item \textbf{Q3}: Compared to the more commonly used mutual correlation, why is combination model IC a better metric?
    \item \textbf{Q4}: How does our framework perform under more realistic trading settings?
\end{itemize}

\subsection{Experiment Settings}

\subsubsection{Data} \label{sec:data}

Our experiments are conducted on raw data from the Chinese A-shares market\footnote{The stock price/volume data is retrieved from \url{https://www.baostock.com}. Regarding dividend adjustment, the price/volume data are all forward-dividend-adjusted respected to the adjustment factors on 2023/01/15.}. We select 6 raw features as the inputs to our alphas: \{open, close, high, low, volume, vwap (volume-weighted average price)\}. The target is set to be the 20-day return of the stocks, selling/buying at the closing price ($\textrm{Ref}(\textrm{close}, -20) / \textrm{close} - 1$). The dataset is split by date into a training set (2009/01/01 to 2018/12/31), a validation set (2019/01/01 to 2019/12/31), and a test set (2020/01/01 to 2021/12/31). In the following experiments, we will use the constituent stocks of the CSI300 and the CSI500 indices of China A-shares as the stock set.

\subsubsection{Compared Methods}

To evaluate how well our framework performs against traditional formulaic alpha generation approaches, we implemented two methods that are designed to generate one alpha at a time. \textbf{GP} is a genetic programming model using the alpha's IC as the fitness measure to generate expression trees. This model is implemented upon the gplearn\footnote{\url{https://github.com/trevorstephens/gplearn}} framework. \textbf{PPO} is a reinforcement learning method, based on the same PPO \cite{ppo} algorithm and expression generator, and uses the single alpha's IC as the episode return instead of the combined performance used in our full framework.

Since only using the top-most alpha to evaluate the frameworks are extremely prone to overfitting on the training data, we also constructed alpha sets with the ones generated by the two single alpha generators. The same combination model is then applied to these alpha sets. Note that the generators still emit alphas in a one-by-one manner, and are agnostic to the combination model's performance. The first method to construct the set (\textbf{top}) is to simply select the top-$k$ alphas emitted by the generator with the highest IC on the training set. The second method (\textbf{filter}) is to select the top-$k$ performing alphas with a constraint that any pair of alpha from the set must not have a mutual IC higher than 0.7.

To better evaluate the model performance, we also compared our approach to several end-to-end machine learning models implemented in the open-source library Qlib \cite{qlib}. The models receive 60 days' worth of raw features as the input, and are trained to predict the 20-day returns directly. Note that these models do not generate formulaic alphas. The hyperparameters of these models are set according to the benchmarks given by Qlib.

\begin{itemize}
    \item \textbf{XGBoost} \cite{xgb} is an efficient implementation of gradient boosting algorithms, which ensembles decision trees to predict stock trends directly.
    \item \textbf{LightGBM} \cite{lgbm} is another popular implementation of gradient boosting.
    \item \textbf{MLP}: A multilayer perceptron (MLP) is a type of fully-con\-nec\-ted feedforward artificial neural network.
\end{itemize}

To demonstrate the effect caused by stochasticity in the training process, each experimental combination with an indeterministic training process is evaluated with 10 different random seeds.

\subsubsection{Evaluation Metrics}

We choose two metrics to measure the performance of our models as follows.

\begin{itemize}
    \item \textbf{IC}, the Pearson's correlation coefficient shown in Eq.~\ref{eq:pearson}.
    \item \textbf{Rank IC}, the rank information coefficient. The rank IC tells how much the ranks of our alpha values are correlated with the ranks of future returns. Rank IC is defined by replacing Pearson's correlation coefficient with Spearman's correlation coefficient. The rank IC is just the IC of ranked data, defined as follows:
          \begin{equation}
              \sigma^{\mathrm{rank}}(u, v) = \sigma(r(u), r(v)),
          \end{equation}
          where $r(\cdot)$ is the ranking operator. The ranks of repeated values are assigned as the average ranks that they would have been assigned to\footnote{For example, $r(\parens{3, -2, 6, 4}) = \parens{2, 1, 4, 3}$, while $r((3, -2, 4, 4)) = \parens{2, 1, 3.5, 3.5}$.}.
\end{itemize}

Both of the metrics are \textbf{the higher the better}.

\begin{table}[]
    \centering
    \caption{Main results on CSI 300 and CSI 500. Values outside parentheses are the means, and values inside parentheses are the standard deviations across 10 runs.}
    \resizebox{0.48\textwidth}{!}{
        \begin{tabular}{p{0.09\textwidth} | >{\centering\arraybackslash}p{0.085\textwidth} >{\centering\arraybackslash}p{0.085\textwidth} | >{\centering\arraybackslash}p{0.085\textwidth} >{\centering\arraybackslash}p{0.085\textwidth}}
            \toprule[1.5pt]
            \multirow{2}{*}{\textbf{Method}} & \multicolumn{2}{c|}{\textbf{CSI 300}} & \multicolumn{2}{c}{\textbf{CSI 500}} \\
            \cmidrule{2-5} & \textbf{IC($\uparrow$)} & \textbf{Rank IC($\uparrow$)} & \textbf{IC($\uparrow$)} & \textbf{Rank IC($\uparrow$)} \\
            \midrule[1pt]
            \multirow{2}{*}{MLP} & $\hphantom{-}0.0250$ & $\hphantom{-}0.0401$ & $0.0188$ & $0.0458$ \\
            & $\hphantom{-}(0.0068)$ & $\hphantom{-}(0.0081)$ & $(0.0018)$ & $(0.0045)$ \\
            XGBoost & $\hphantom{-}0.0404$ & $\hphantom{-}0.0576$ & $0.0353$ & $0.0639$ \\
            LightGBM & $\hphantom{-}0.0259$ & $\hphantom{-}0.0324$ & $0.0332$ & $0.0609$ \\

            \hline

            \multirow{2}{*}{PPO\_top\textbf{*}} & $-0.0166$ & $-0.0144$ & $0.0025$ & $0.0295$ \\
            & $\hphantom{-}(0.0028)$ & $\hphantom{-}(0.0075)$ & $(0.0076)$ & $(0.0135)$ \\
            \multirow{2}{*}{GP\_top\textbf{*}} & $\hphantom{-}0.0078$ & $\hphantom{-}0.0157$ & $0.0200$ & $0.0504$ \\
            & $\hphantom{-}(0.0218)$ & $\hphantom{-}(0.0271)$ & $(0.0112)$ & $(0.0160)$ \\
            \multirow{2}{*}{PPO\_filter\textbf{*}} & $-0.0044$ & $\hphantom{-}0.0101$ & $0.0042$ & $0.0506$ \\
            & $\hphantom{-}(0.0107)$ & $\hphantom{-}(0.0107)$ & $(0.0042)$ & $(0.0052)$ \\
            \multirow{2}{*}{GP\_filter\textbf{*}} & $\hphantom{-}0.0183$ & $\hphantom{-}0.0298$ & $0.0117$ & $0.0562$ \\
            & $\hphantom{-}(0.0190)$ & $\hphantom{-}(0.0227)$ & $(0.0083)$ & $(0.0105)$ \\
            \hline
            \multirow{2}{*}{\textbf{Ours\textbf{*}}} & $\hphantom{-}\textbf{0.0725}$ & $\hphantom{-}\textbf{0.0806}$ & $\textbf{0.0438}$ & $\textbf{0.0727}$ \\
            & $\hphantom{-}(0.0105)$ & $\hphantom{-}(0.0106)$ & $(0.0064)$ & $(0.0112)$ \\

            \bottomrule[1.5pt]
            \multicolumn{5}{c}{\textbf{*}Optimal combination size in $\braces{10, 20, 50, 100}$} \\
        \end{tabular}}
    \label{tab:main_ic_rankic}
\end{table}

\begin{figure}[]
    \centering
    \makebox[\linewidth][c]{\includegraphics[width=1.08\linewidth]{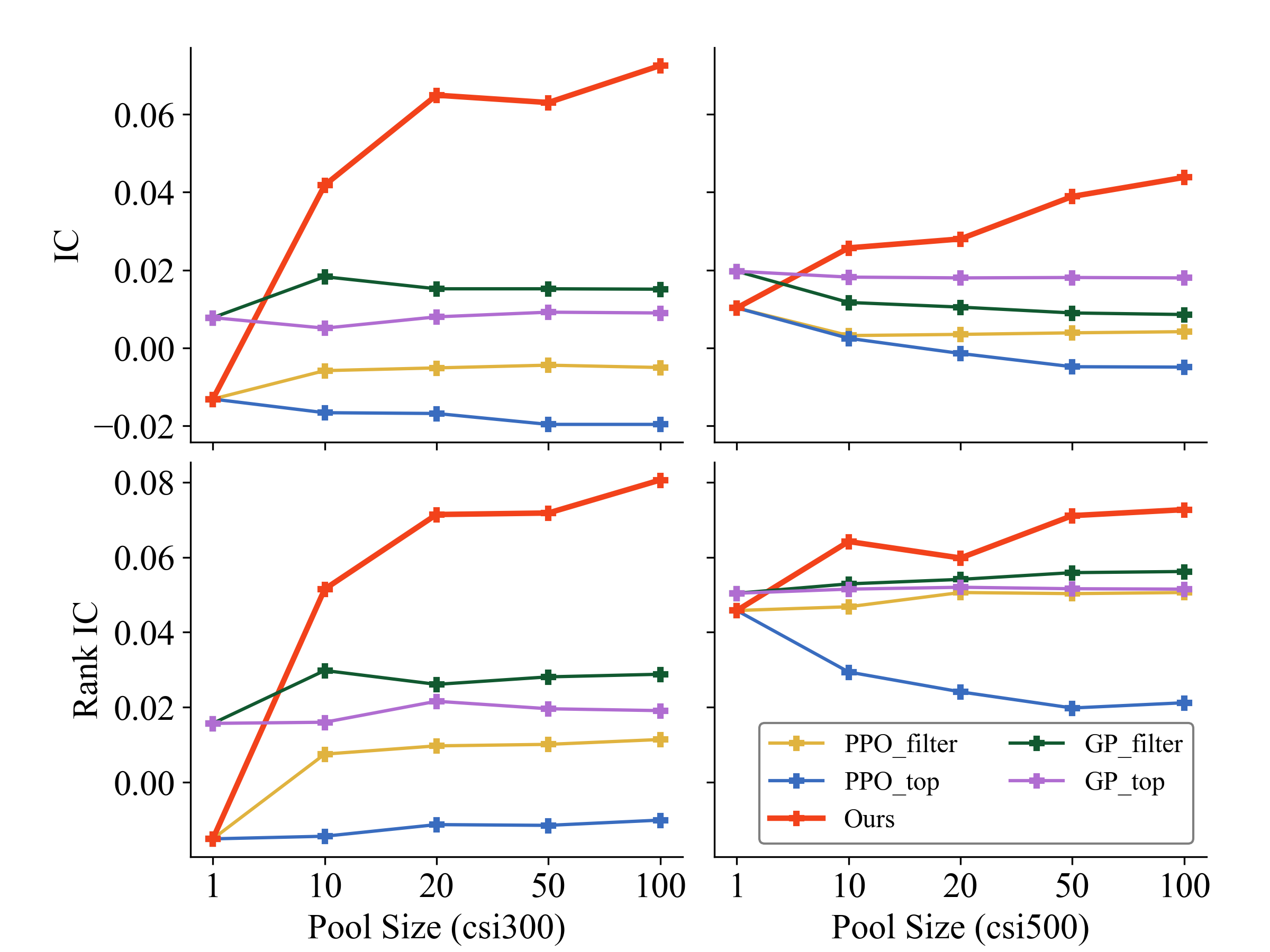}}
    \caption{The results of ablation study. A pool size of 1 refers to settings that only evaluate the top-most alpha without using a combination model.}
    \label{fig:ablation}
\end{figure}

\begin{table*}[ht]
    \centering
    \caption{An example combination of 10 alphas.}
    \begin{tabular}{>{\centering\arraybackslash}p{0.03\textwidth} | >{\centering\arraybackslash}p{0.6\textwidth} | >{\centering\arraybackslash}p{0.14\textwidth} >{\centering\arraybackslash}p{0.14\textwidth}}
        \toprule[1.5pt]
        \# & Alpha & Weight & IC (CSI300) \\
        \midrule[1pt] 1 & $\mathrm{Var}(\mathrm{Greater}(\mathrm{Greater}(\mathrm{Var}(\mathrm{low},50),\mathrm{high}),\mathrm{open}),30)$
        & $-0.0295$ & \cellcolor{p0000}$\hphantom{-}0.0011$ \\
        2 & $\mathrm{Max}((\mathrm{Min}(\mathrm{Max}(\mathrm{close}, 20) - 30, 20) + 100) / 30, 20)$
        & $\hphantom{-}0.0515$ & \cellcolor{p0025}$\hphantom{-}0.0262$ \\
        3 & $(\mathrm{Mad}(\mathrm{high}, 50) + 0.5)\mathrm{vwap} / \mathrm{close}$
        & $\hphantom{-}0.0343$ & \cellcolor{p0040}$\hphantom{-}0.0447$ \\
        4 & $\mathrm{Ref}(\mathrm{low}, 50)$
        & $\hphantom{-}0.0260$ & \cellcolor{p0020}$\hphantom{-}0.0241$ \\
        5 & $\mathrm{Min}(\mathrm{high} / \mathrm{close}, 50) - \mathrm{Greater}(-0.05 / \mathrm{close}, -10)$
        & $\hphantom{-}0.0437$ & \cellcolor{n0025}$-0.0211$ \\
        6 & $\mathrm{Delta}(\mathrm{high}, 20) + \mathrm{high} - 12$
        & $-0.0997$ & \cellcolor{p0015}$\hphantom{-}0.0165$ \\
        7 & $\mathrm{Less}(\mathrm{Min}(2(\mathrm{vwap} - \mathrm{volume}), 30) + 30, 30 \mathrm{vwap} / \mathrm{low}) - 5$
        & $\hphantom{-}0.0276$ & \cellcolor{p0000}$\hphantom{-}0.0025$ \\
        8 & $\mathrm{Corr}(\mathrm{Greater}(\mathrm{Greater}(\mathrm{vwap}, \mathrm{volume}),\mathrm{Greater}(\mathrm{close},$ \newline $\mathrm{Greater}(\mathrm{Log}(\mathrm{Var}(\mathrm{volume},10)),10))/\mathrm{close}),\mathrm{close},10)$
        & $-0.0279$ & \cellcolor{n0035}$-0.0338$ \\
        9 & $|\mathrm{low} - 30|$
        & $\hphantom{-}0.0319$ & \cellcolor{p0005}$\hphantom{-}0.0073$ \\
        10 & $\mathrm{Max}(1 - \mathrm{Max}(\mathrm{Corr}(\mathrm{low}, \mathrm{volume} \cdot \mathrm{Log}(10^{-4}\mathrm{Max}(\mathrm{volume}, 10)), 10), 10), 30)$
        & $\hphantom{-}0.0312$ & \cellcolor{p0045}$\hphantom{-}0.0488$ \\
        \hline
        \multicolumn{3}{c}{Weighted Combination} & \cellcolor{p0050}$\hphantom{-}0.0511$ \\
        \bottomrule[1.5pt]
    \end{tabular}
    \label{tab:case_study}
\end{table*}

\subsection{Main Results}

\subsubsection{Comparison across all alpha generators}

To answer \textbf{Q1}, we first compare our framework against several other alpha-mining methods and direct stock trend forecasting baselines, including PPO, GP, MLP, LightGBM, and XGBoost. Experiments are conducted on CSI300 and CSI500 stocks respectively.

The results are shown in Table \ref{tab:main_ic_rankic}. Our framework is able to achieve the highest IC and rank IC across all the methods we compare to. Note that the framework is only explicitly optimized against the IC metric. The non-formulaic alpha models come in the second tier. The baseline formulaic alpha generators perform poorly on the test set, especially the RL-based ones. The reinforcement learning agent, when optimized only against single-alpha IC, is prone to falling into local optima and thus overfitting on the training set, and basically stops searching for new alphas after a certain amount of steps. On the other hand, the GP-based methods maintaining a large population can avoid the same problem, but still cannot produce alphas that are synergistic when used together. The results also show that the filtering techniques cannot solve the synergy problem consistently either.

\subsubsection{Comparison of formulaic generators with varying pool capacity} \label{sec:ablation}

To answer \textbf{Q2}, we study the four baseline formulaic alpha generators more extensively, and compare them to our proposed framework. The models are evaluated under pool sizes of $k \in \braces{1, 10, 20, 50, 100}$. The results are shown in Figure \ref{fig:ablation}.

Compared to the baseline method PPO\_filter, our method directly uses the combination model's performance as the reward to newly generated alphas. This leads to a substantial improvement when the pool size increases, meaning that our method can produce alpha sets with great synergy. Our method shows scalability for pool size: even when the pool size is large enough, it can still continuously find synergistic alphas that boost the performance over the existing pool. Conversely, the combined performance of the alphas generated by other approaches barely improves upon the case with just the top alpha, meaning that these alpha factors have poor synergy. Furthermore, the ability to control the reward of individual expressions under a certain alpha pool configuration is granted by the flexibility of the RL scheme. The GP scheme of maintaining a large population at the same time does not work well with fine-grained fitness value control.

Also, we can see that for the CSI500 dataset, GP\_filter performs worse than GP\_top on the IC metric when the pool size increases. This phenomenon demonstrates that the traditionally used mutual-IC filtering is not always effective, answering the question \textbf{Q3}.

\subsection{Case Study}

Table \ref{tab:case_study} shows an example combination of 10 alphas generated by our framework, evaluated on the CSI300 constituent stock set. Most of the alpha pairs in this specific set have mutual IC values over 0.7. Previous work \cite{huataigp}\cite{autoalpha} considered this to be too high for the individual alphas to be regarded as ``diverse'', yet these alphas are able to work well in a synergistic manner. For example, the alphas \#2 and \#6 have a mutual IC of $0.9746$, thus traditionally considered too similar to be useful cooperatively. However, the combination $0.09317f_2 - 0.07163f_6$ achieves an IC of $0.0458$ on the test set, even higher than the sum of the respective ICs, showing the synergy effect.

Also, although alpha \#1 only has an IC of 0.0011, it still plays a vital role in the final combination. Once we remove alpha \#1 from the combination and re-train the combination weights on the remaining set, the combination's IC drops to merely $0.0447$. The two observations above show that neither the single alpha IC nor the mutual IC between alpha pairs is a good indicator of how well the combined alpha would perform, answering \textbf{Q3}.

One possible explanation for these phenomena is that: Although traditionally these alphas are similar due to the high mutual IC, some linear combinations of the alphas could point to a completely different direction from the original ones. Consider two unit vectors in a linear space. The more similar these two vectors are, the less similar either of these vectors is to the difference between the two vectors, since the difference vector approaches to be perpendicular to either of the original vectors as the vectors get closer.\

\begin{figure*}[]
    \centering
    \includegraphics[width=1.0\linewidth]{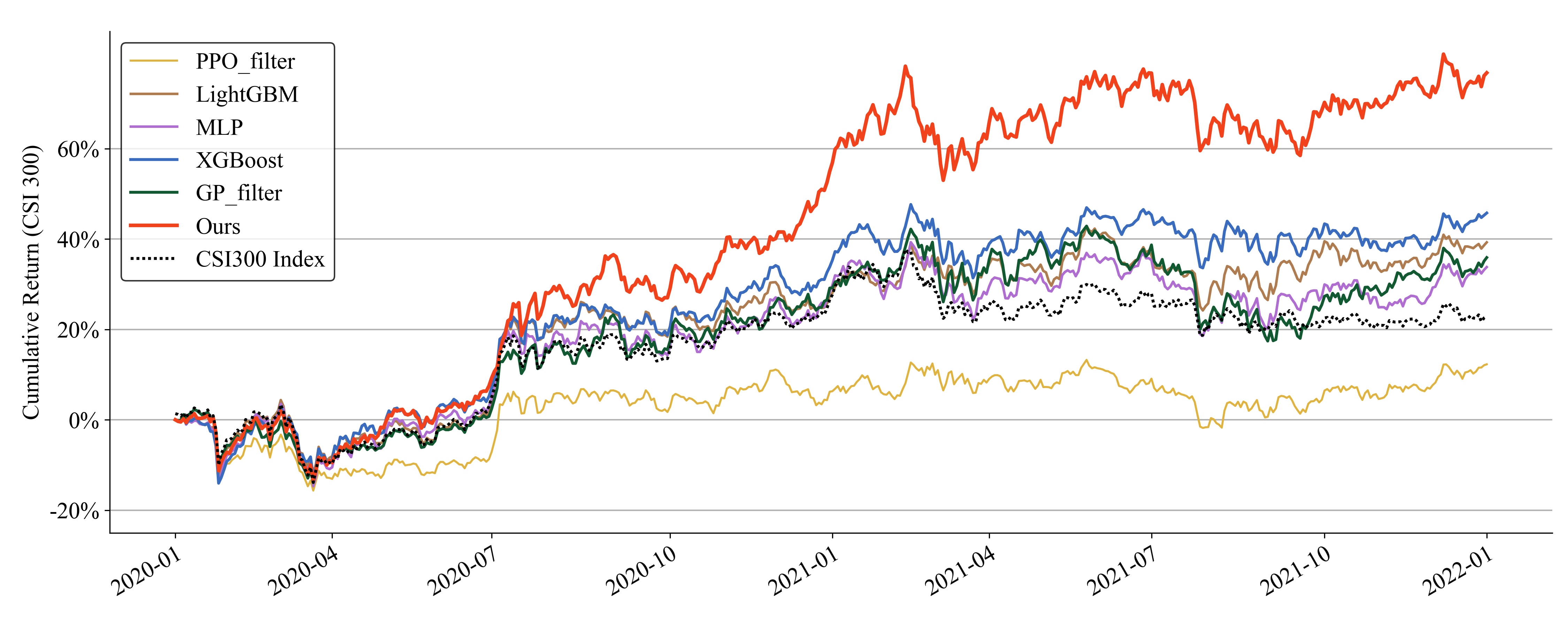}
    \caption{Backtest results on CSI 300. The lines track the net worth of simulated trading agents utilizing the various alpha-mining approaches.}
    \label{fig:backtest}
\end{figure*}

\subsection{Investment Simulation}

To demonstrate the effectiveness of our factors in more realistic investing settings, we use a simple investment strategy and conducted backtests in the testing period (2020/01/01 to 2021/12/31) on the CSI300 dataset. We use a simple \emph{top-$k$/drop-$n$} strategy to simulate the investment: On each trading day, we first sort the alpha values of the stocks, and then select the top $k$ stocks in that sorted list. We evenly invest across the $k$ stocks if possible, but restrict the strategy to only buy/sell at most $n$ stocks on each day to reduce excessive trading costs. In our experiment, $k$ is set to 50 and $n$ to 5.

We recorded the net worth of the respective strategies in the testing period, of which a line chart is shown in Figure \ref{fig:backtest}. Although our framework does not explicitly optimize towards the absolute returns, the framework still performs well in the backtest. Our framework is able to gain the most profit compared to the other methods.

\section{Related Work}\label{sec:relatedwork}

\textbf{Formulaic alphas.} The search space of formulaic alphas is enormous, due to the large amount of possible operators and features to choose from. To our best knowledge, all notable former work uses genetic programming to explore this huge search space. \cite{huataigp} augmented the gplearn library with formulaic-alpha-specific time-series operators, upon which an alpha-mining framework is built. \cite{huataigp2} further improved the framework to also mine alphas with non-linear relations with the returns by using mutual information as the fitness measure. \cite{autoalpha} used mutual IC to filter out alphas that are too similar to existing ones, improving the diversity of resulting alpha sets. PCA is carried out on the alpha values for reducing the algorithmic complexity of computing the mutual ICs, and various other tricks are also applied to aid the evolution process. AlphaEvolve \cite{alphaevolve} evolves new alphas upon existing ones. It allows combinations of much more complex operations (for example matrix-wise computations), and uses computation graphs instead of trees to represent the alphas. This leads to more sophisticated alphas and better prediction accuracy, although at the risk of lowering the alphas' interpretability. Mutual IC is also used as a measure of alpha synergy in this work.

\textbf{Machine learning-based alphas.} The development of deep learning in recent years has brought about various new ideas on how to accurately model stock trends. Early work on stock trend forecasting treats the movement of each stock as a separate time series, and applies time series models like LSTM \cite{lstm} or Transformer \cite{transformer} to the data. Specific network structures catered to stock forecasting like the SFM \cite{sfm} which uses a DFT-like mechanism have also been developed. Recently, research has also been conducted on methods to integrate non-standard data with the time series. REST \cite{rest} fuses multi-granular time series data together with historical event data to model the market as a whole. HIST \cite{hist} utilizes concept graphs on top of the regular time series data to model shared commonness between future trends of various stock groups. One specific type of machine learning-based model is also worth mentioning. Decision tree models, notably XGBoost \cite{xgb}, LightGBM \cite{lgbm}, etc., are often considered interpretable, and they could also achieve relatively good performance on stock trend forecasting tasks. However, whether a decision tree with extremely complex structure is considered ``interpretable'' is at least questionable. When these tree models are applied to raw stock data, the high dimensionality of input only exacerbates the aforementioned problem. Our formulaic alphas use operators that apply to the input data in a more structured manner, making them more easily interpretable by curious investors.

\textbf{Symbolic regression.} Symbolic regression (SR) concerns the problem of discovering relations between variables represented in closed-form mathematical formulas. SR problems are different from our problem settings that there always exists a ``groundtruth`` formula that precisely describes the data points in an SR problem, while stock market trends are far too complex to be expressed in the space of formulaic alphas. Nevertheless, there remain similarities between the two fields since similar techniques can be used for the expression generator and the optimization procedure. \cite{srnn} suggested using a custom neural network whose activation functions are symbolic operators to solve the SR problem. \cite{dsr} proposed a novel symbolic regression framework based on an autoregressive expression generator. The generator is optimized using an augmented version of the policy gradient algorithm that values the top performance of the agent more than the average. \cite{dsrgp} developed a method similar to \cite{dsr}, but also introduced GP into the optimization loop, seeding the GP population with RL outputs. \cite{symbolicgpt} applied the language model pretraining scheme to symbolic regression, training a generative autoregressive ``language model'' of expressions on a large dataset of synthetic expressions.

\textbf{Discussions.} Although the term ``formulaic alpha'' is often tied down to investing, the concept of simple and interpretable formulaic predictors that could be combined into more expressive models is not limited to quantitative trading scenarios. Our framework can be adapted to solve other time-series forecasting problems, for example, energy consumption prediction \cite{disc1}, anomaly detection \cite{disc2}, biomedical settings \cite{disc3}, etc. In addition, we chose the linear combination model in this paper for its simplicity. Meanwhile, in theory, other types of interpretable combination models, for example, decision trees can also be integrated into our framework. In that sense, providing these combination models with these features expressed in relatively straightforward formulas might help provide investigators with more insights into how the models come to the final results.

\section{Conclusion}\label{sec:conclusion}

In this paper, we proposed a new framework for generating interpretable formulaic alphas to aid investors in quantitative trading. We proposed to directly use the performance boost brought about by the newly added alpha to the existing alpha combination as the metric for alpha synergy. As a result, our framework can produce sets of alphas that could cooperate satisfactorily with a combination model, notwithstanding the actual form of the combination model. For the model to explore the vast search space of formulaic alphas more effectively, we also formulated the alpha-searching procedure as an MDP and applied reinforcement learning techniques to optimize the alpha generator. Extensive experiments are conducted to demonstrate that the performance of our framework surpasses those of all previous formulaic alpha-mining approaches, and that our method can also perform well under more realistic trading settings.

\begin{acks}

The research work was supported by the National Key Research and Development Program of China under Grant No. 2022YFC3303302, the National Natural Science Foundation of China under Grant No.61976204. Xiang Ao is also supported by the Project of Youth Innovation Promotion Association CAS, Beijing Nova Program Z201100006820062.

\end{acks}

\FloatBarrier

\bibliographystyle{ACM-Reference-Format}
\balance
\bibliography{references}
\appendix
\begin{table*}[]
	\centering
	\caption{All the operators used in our framework. CS: cross-section, TS: time-series, U: unary, B: binary.}
    \begin{tabularx}{\textwidth}{>{\hsize=.4\hsize}XcX}
        \toprule[1.5pt]
        Operator & Category & Descriptions \\
        \midrule
        Abs($x$) & CS--U & The absolute value $|x|$. \\
        Log($x$) & CS--U & Natural logarithmic function $\log(x)$. \\
        \hline
        $x+y$, $x-y$, $x \cdot y$, $x/y$ & CS--B & Arithmetic operators. \\
        $\mathrm{Greater}(x, y)$, $\mathrm{Less}(x, y)$ & CS--B & The larger/smaller one of the two values. \\
        \hline
        $\mathrm{Ref}(x, t)$ & TS--U & The expression $x$ evaluated at $t$ days before the current day. \\
        $\mathrm{Mean}(x, t)$, $\mathrm{Med}(x, t)$, $\mathrm{Sum}(x, t)$ & TS--U & The mean/median/sum value of the expression $x$ evaluated on the recent $t$ days. \\
        $\mathrm{Std}(x, t)$, $\mathrm{Var}(x, t)$ & TS--U & The standard deviation/variance of the expression $x$ evaluated on recent $t$ days. \\
        $\mathrm{Max}(x, t)$, $\mathrm{Min}(x, t)$ & TS--U & The maximum/minimum value of the expression $x$ evaluated on the recent $t$ days. \\
        $\mathrm{Mad}(x, t)$ & TS--U & The mean absolute deviation $\mathbb{E}\brackets{|x - \mathbb{E}\brackets{x}|}$ of the expression $x$ evaluated on the recent $t$ days. \\
        $\mathrm{Delta}(x, t)$ & TS--U & The relative difference of $x$ compared to $t$ days ago, $x - \mathrm{Ref}(x, t)$. \\
        $\mathrm{WMA}(x, t)$, $\mathrm{EMA}(x, t)$ & TS--U & Weighted moving average and exponential moving average of the expression $x$ evaluated on the recent $t$ days. \\
        \hline
        $\mathrm{Cov}(x, y, t)$ & TS--B & The covariance between two time series $x$ and $y$ in the recent $t$ days. \\
        $\mathrm{Corr}(x, y, t)$ & TS--B & The Pearson's correlation coefficient between two time series $x$ and $y$ in recent $t$ days. \\
        \bottomrule[1.5pt]
    \end{tabularx}
	\label{tab:operators}
\end{table*}

\section{List of Operators} \label{sec:ops}

There are four types of operators used in our framework. The four types break down into two groups: cross-section operators, and time-series operators. Cross-section operators (denoted with ``CS'' in the table) only deal with data on the current trading day, while time-series operators (denoted with ``TS'') take into consideration data from a consecutive period of time. Each of the two groups further separates into unary (denoted with ``U'') and binary (denoted with ``B'') operators that apply to one or two series respectively.

\section{Proof of Theorem \ref{thm:loss_transformation}} \label{sec:thm_proof}

\begin{proof}
    
We know that the elements of a vector $u$ that is centralized and normalized (using the $\mathcal{N}$ operator mentioned above) have a variance of $1/n$, since:
\begin{equation}
    \begin{aligned}
        \var\brackets{u}
        & = \expectation{i}{u_i^2} - \expectation{i}{u_i}^2 \\
        & = \frac{1}{n}\|u\|^2 - 0 \\
        & = \frac{1}{n}.
    \end{aligned}
\end{equation}

Using the original definition of Pearson's correlation coefficient, we have:
\begin{equation}
    \begin{aligned}
        \sigma(u, v)
        & = \frac{\cov\brackets{u, v}}{\sqrt{\var\brackets{u} \cdot \var\brackets{v}}} \\
        & = \expectation{i}{
            \frac{(u_i - \bar{u})}{\sqrt{\var\brackets{u}}}
            \cdot
            \frac{(v_i - \bar{v})}{\sqrt{\var\brackets{v}}}
        } \\
        & = \expectation{i}{
            \frac{\brackets{\mathcal{N}(u)}_i}{\sqrt{1/n}}
            \cdot
            \frac{\brackets{\mathcal{N}(v)}_i}{\sqrt{1/n}}
        } \\
        & = n\expectation{i}{\brackets{\mathcal{N}(u)}_i \brackets{\mathcal{N}(v)}_i} \\
        & = \angles{\mathcal{N}(u), \mathcal{N}(v)}.
    \end{aligned}
\end{equation}
That is to say, the Pearson's correlation coefficient between two vectors equals the inner product of the two vectors centralized and normalized.

Therefore the theorem can be proved as follows. Recall that $f_i(x_t)$ and $y_t$ are normalized.
\begin{equation}
    \begin{aligned}
        n\mathcal{L}(w)
        & = \frac{1}{T} \sum_{t=1}^T \|z_t - y_t\|_2^2 \\
        & = \expectation{t}{\|z_t\|^2 - 2\angles{z_t, y_t} + \|y_t\|^2} \\
        & = \expectation{t}{
            \Verts{\sum_{i=1}^k w_if_i(X_t)}^2
            - 2 \angles{\sum_{i=1}^k w_if_i(X_t), y_t}
            + 1
        } \\
        & = \mathbb{E}_t\left[ \sum_{i=1}^k\sum_{j=1}^k w_iw_j \sigma(f_i(X_t), f_j(X_t))\right. \\
        & \qquad \left. -\ 2\sum_{i=1}^k w_i\angles{f_i(X_t), y_t} + 1 \right] \\
        & = \sum_{i=1}^k\sum_{j=1}^k w_iw_j \bar{\sigma}(f_i(X), f_j(X)) - 2\sum_{i=1}^k w_i \bar{\sigma}_y(f_i) + 1.
    \end{aligned}
\end{equation}

\end{proof}

\section{Expression Legality Guarantee} \label{sec:expr_legality}

The legality of expressions divides into two parts: \emph{Formal} legality and \emph{semantic} legality.

\subsection{Formal Legality}

An RPN can be built with a stack of expressions, constants, or raw features. The RPN building procedure follows the rules below, and actions that may violate these rules will be masked. 

\begin{itemize}
    \item TS (time-series) operators must take a time-delta (e.g. 10d for a time-difference of 10 days) as its last parameter;
    \item Excluding the aforementioned time-delta, each operator must take enough expressions as operands, according to the arity of the operator (one for *-Unary, two for *-Binary);
    \item A multi-token expression should not be equivalent to a constant;
    \item The special SEP token (end of expression) is only allowed when the generated sequence is already a valid RPN.
\end{itemize}

For example, when the stack (state) is currently [\$open, 0.5], we can choose the ``Add'' token (a binary operator), building an expression ``Add(\$open, 0.5)''. Meanwhile, the operator ``Log'' is not allowed here because ``Log'' will take ``0.5'' and ``Log(0.5)'' is a constant; similarly, the operator ``TS-Mean''
is also invalid because ``Mean(\$open, 0.5)'' is illegal.

\subsection{Semantic Legality}

Some expressions with correct forms might still fail to evaluate due to more constraints imposed by the operators. For example, the logarithm operator cannot be applied to a non-positive value. This kind of semantic invalidity is not directly detected by the procedure mentioned in the last section. In our experiments, these expressions are given the reward of -1 (the minimum value of Pearson's correlation coefficient) to discourage the agent from generating these expressions.

\section{Hyperparameters}
\label{sec:hyperparam}

The LSTM feature extractor used in the RL agent has a 2-layer structure with a hidden layer dimension of 128. A dropout rate of 0.1 is used in the LSTM network. The separate value and policy heads are MLPs with two hidden layers of 64 dimensions. PPO clipping range $\epsilon$ is set to 0.2.

\end{document}